\documentclass[twocolumn]{emulateapj}
\newcommand{\mpc}{\rm {h^{-1}Mpc }}

\newcommand{\avg}[1]{\langle{#1}\rangle}

\newcommand{\ltsima}{$\; \buildrel < \over \sim \;$}
\newcommand{\lsim}{\lower.5ex\hbox{\ltsima}}
\newcommand{\gtsima}{$\; \buildrel > \over \sim \;$}
\newcommand{\gsim}{\lower.5ex\hbox{\gtsima}}

\def\gtrsim{\mathrel{\hbox{\rlap{\hbox{\lower4pt\hbox{$\sim$}}}\hbox{$>$}}}}

\def\lesssim{\mathrel{\hbox{\rlap{\hbox{\lower4pt\hbox{$\sim$}}}\hbox{$<$}}}}

\begin{document}
\title{Fast Edge Corrected Measurement of the Two-Point Correlation 
Function and the Power Spectrum}

\author{Istv\'an Szapudi\altaffilmark{1}, Jun Pan\altaffilmark{1,2}, 
Simon Prunet\altaffilmark{3}, and Tam\'as Budav\'ari\altaffilmark{4}}
 
\altaffiltext{1}{Institute for Astronomy, University of Hawaii,
2680 Woodlawn Dr, Honolulu, HI 96822, USA}
\altaffiltext{2}{The School of Physics and Astronomy,
University of Nottingham, University Park, Nottingham, NG7 2RD, England}
\altaffiltext{3}{Institut d'Astrophysique de Paris, CNRS,
98bis bd Arago, F-75014 Paris, France}
\altaffiltext{4}{Dept. of Physics and Astronomy, The Johns Hopkins University
3400 North Charles Street, Baltimore, MD 21218-2686}

\begin{abstract}

We present two related techniques to measure the
two-point correlation function and the power spectrum
with edge correction in any spatial dimensions.
The underlying algorithm uses fast Fourier transforms for calculating the
two-point function with an heuristically weighted 
edge corrected estimator. Once the correlation function is measured,
we estimate the power spectrum via numerical integration of
the Hankel transform connecting the two. We introduce an efficient numerical
technique based on Gauss-Bessel-quadrature and double
exponential transformation. This, combined with our, or any
similar, two-point function estimator leads to a novel edge corrected 
estimator for power spectra. The pair of techniques presented
are the Euclidean analogs of those developed and widely used
in cosmic microwave background research for spherical maps.

\end{abstract}


\section{Introduction}

The two-point correlation function and its Fourier transform,
the power spectrum, are the most measured statistics in 
large scale structure studies. Correlation functions 
can be measured naively with an $O(N^2)$ technique through
cycling through pairs of objects. The advantage of
that is that the edge corrected estimator of \cite{LandySzalay1993}
or  \cite{SzapudiSzalay1998,SzapudiSzalay2000} 
(c.f,  Equation~\ref{eq:estimator}) can be used for nearly
optimal measurements. The naive counting of pairs can
be sped up, if one is interested in measuring the
correlation function on small scales, and reasonably
large bins. In that case the double-tree algorithm by \cite{MooreEtal2001}
scales approximately as $O(N\log N)$. However, if all
scales are considered, with high resolution, the tree-based
algorithm has to split all nodes of the tree, thus it will
degenerate into the $O(N^2)$ scaling of the naive algorithm.

Measurement of the power spectrum can be obtained
directly with fast Fourier transforms (FFTs).
The scaling of this algorithm $O(N\log N)$,
where $N$ corresponds to the dynamic range of the measurement.
A particular advantage of the FFT is that the speed is independent of
the maximum scale (or smallest wavenumber): scales up to the
fundamental mode can be measured together with scales corresponding
to the grid size.
The resulting estimates, however,  are convolved with the survey
window. In cosmic microwave background (CMB) studies, such power
spectra are termed pseudo-power spectra, and have been entirely
replaced by edge corrected (or true) power spectra 
(\citep[e.g.,][]{SzapudiEtal2001b,HivonEtal2002}). In this
paper we ask the question: can we take advantage of the
FFT algorithms, and calculate the correlation function
in $O(N\log N)$ time even on large scales? Can we exploit the edge
corrected estimator for the two-point correlation function,
and estimate true (edge-corrected) power spectra? As
we show next, the answer is affirmative to both questions,
in complete analogy to results already obtained in CMB research.

In the next \S 2  we describe a new technique to measure
edge corrected correlation functions using an
FFT based algorithm, and a numerical technique
to obtain edge corrected power spectra. We present
numerical results in \S 3, and summarize and discuss them
in the last \S 4.

\section{Description of the Technique}

It is well known that the power spectrum is the Fourier transform
of the two-point correlation function (Wiener-Khinchin theorem).
In cosmic microwave background research, this basic idea was exploited
by \cite{SzapudiEtal2001b} to obtain a fast method for estimating
the angular power spectrum $C_\ell$'s with fast harmonic transforms.
A crucial intermediate step in their method was to estimate correlation
functions \citep[][]{SzapudiEtal2001a}. Although the
window function can be inverted in harmonic space
\citep{HivonEtal2002}, edge effect correction can be trivially obtained from
the estimator \cite{SzapudiSzalay1998,SzapudiSzalay2000}. This
correspond to inverting the window matrix in pixel space, where it is 
diagonal.
Our plan is to adapt this method for
Euclidean space, arbitrary dimensions. The resulting technique
is naturally divided into two steps: i) fast calculation of the
two-point correlation function via FFT ii) calculation of the
power spectrum via a numerical Hankel transform.

\subsection{Two-point correlation function}

The minimum variance estimator of \cite{SzapudiSzalay1998} is most natural
when defined on a grid. If $\delta_i$ is the density field 
on grid point $i$, we can define the estimator as

\begin{equation}
 \tilde\xi_\Delta  = \sum_{i,j} f_{ij}\delta_i\delta_j/\sum_{i,j} f_{ij},
\label{eq:estimator}
\end{equation}
where $f_{ij}$ is the pair weight, and we assumed that $\avg{\delta_i} = 0$.
Naive estimation with general pair weight is $O(N^2)$, where $N$ is
the number of pixels. As we will see, fast estimation is possible
in the special case of the pair weight, where 
$f_{ij} = f_i f_j \delta^{Kr}_{i-j,\Delta}$, i.e. the weight is 
multiplicative, and depends only on a shift $\Delta$. Here $\delta^{Kr}_{kl}$
is the Kronecker-$\delta$, taking values of $1$ when $k=l$, and $0$
otherwise. The simplest example is flat weighting $f_i=1$. 

The correlation function without the normalization
is a ``raw-correlation'' function of unnormalized counts.
The normalization for an arbitrary complex geometry can be obtained
by calculating the raw correlation function of the geometry, 
putting $1$'s into valid grid points, and $0$'s everywhere else.
While our notation suggests $D=1$ dimensions,
in fact arbitrary dimension is included, simply
by replacing indices, say $i$, with tuples of indices $(i_1,\ldots,i_D)$.

Computationally, one needs to calculate
pair summations of the form $\sum a_i a_{i+\Delta}$.
These  be reformulated to make use of
Fast Fourier Transforms (FFTs), one of fastest algorithms in existence.
If $P(x) = a_0 + a_1 x + \ldots + a_{n-1} x^{n-1}$ is a polynomial,
and $\epsilon = e^{2\pi i/r}$ unit roots, the coefficients
of a discrete Fourier transform of the series $a_i$ can be defined as
\begin{equation}
  \hat a_k = P(\epsilon^k) = a_0+a_1\epsilon^k+\ldots+a_{n-1}\epsilon^{(n-1)k}.
\end{equation}
Direct calculation confirms that $\sum a_k b_{k+\Delta}$ can
be calculated by Fourier transforming the series $a_i$ and $b_i$,
multiplying the resulting $\hat a_k \hat b_k^*$, and finally inverse
Fourier transforming back. This simple observation is the discrete
analog of the Wiener-Khinchin theorem. The Fourier space
quantity arising in the above algorithm is the discrete,
Euclidean ``pseudo-power spectrum''. Note that in fact
all previous direct measurements of the power spectrum
using \cite{FeldmanEtal1994} type estimators measured
pseudo power spectrum. We will see in the next subsection,
how to obtain the (edge corrected) power spectrum instead
from the correlation function.

Using the above observation, we can put together a fast 
algorithm to calculate two-point correlation function on
in  $D$ dimensions with the following steps: i) placing the points in a
sufficiently fine grid, storing the value $N_{\bf k}$, the number
of objects (e.g. galaxies) at (vector) grid point ${\bf k}$, 
(this step is omitted
if the density field is a given, such as Euclidean approximation
of CMB maps, etc.)
ii) calculating fluctuations of the field by
$\delta = (N-\avg{N})/\avg{N}$ for each grid point iii) 
possible weighting each point with a minimum variance
weight (e.g. $J_3$ weighting, or \cite{FeldmanEtal1994} weighting), iv)
discrete Fourier transform with a fast FFT engine, 
v) multiplying the coefficients vi)
Fourier transform back. The same procedure is followed for
the mask with zero padding large enough (i.e. larger than
the largest bin of the correlation function) to avoid aliasing effects.
Finally, the raw correlation function is divided with the
correlation function of the mask, according to Equation~\ref{eq:estimator}.

The result will be an inhomogeneous correlation
function $\tilde\xi({\bf r})$, depending on vector shifts ${\bf r}$.
To obtain the traditional correlation function one has sum
the result in rings/spheres for $D=2,3$, respectively.

\subsection{Power Spectrum}

The power spectrum is a $D$-dimensional Fourier transform
of the two-point correlation function. Using rotational
invariance, this will become a Hankel transform in
$D=2,3$ dimensions:
\begin{eqnarray}
   \xi(r)& = \int \frac{k dk}{2\pi}P(k) J_0(kr) \,\, \rm{2D},\cr
   \xi(r)& = \int\frac{k^2 dk}{2\pi^2} P(k) j_0(kr) \,\, \rm{3D},
    \label{eq:2ptpk}
\end{eqnarray}
where $J_0$ and $j_0$ are ordinary and spherical Bessel functions,
respectively. As well known the Hankel transform is self inverse,
and our plan is to use the inverse of the above formulae to
obtain an estimator of the power spectrum. 

In practice, it is a delicate numerical procedure to 
calculate a Hankel transform numerically. We use the quadrature
formulae of \cite{Ogata2005}, which uses both the roots of
Bessel functions, as well as the double exponential transformation
of \cite{OouraMori1999}. Since these techniques are fairly new developments
in numerical mathematics, we summarize the recipe here.
Explicitly, to integrate over an arbitrary function $f$ multiplied
with a Bessel function we use the formula
\begin{eqnarray}
  \int_0^\infty f(x) J_\nu(x) \simeq & \pi\sum_{k=1}^\infty w_{\nu k}
   f(\pi\psi(h r_{\nu k})/h) \times\nonumber\\
  & J_\nu(\pi\psi(h r_{\nu k})/h)\psi^\prime(h r_{\nu k}),
  \label{eq:bessel}
\end{eqnarray}
where ${r_{\nu k}}$ are the roots of the Bessel function $J_\nu$,
$\psi(t) = t\tanh(\frac{\pi}{2}\sinh t)$ the double exponential
transformation, $h$ is the step of the integration (similarly
to that of the trapezoidal rule). The weights are calculated
as $w_{\nu k} = 2/(\pi^2 r_{\nu k} J_{\nu+1}(\pi r_{\nu k}))$.
For the inverting the 
two-point function $\nu=0$ and $\nu=1/2$ Bessel functions 
should be used for 2 and 3 spatial dimensions according to
Eqs~\ref{eq:2ptpk}. 

\section{Results}

\begin{figure}[htb]
\epsscale{1.35}
\plotone{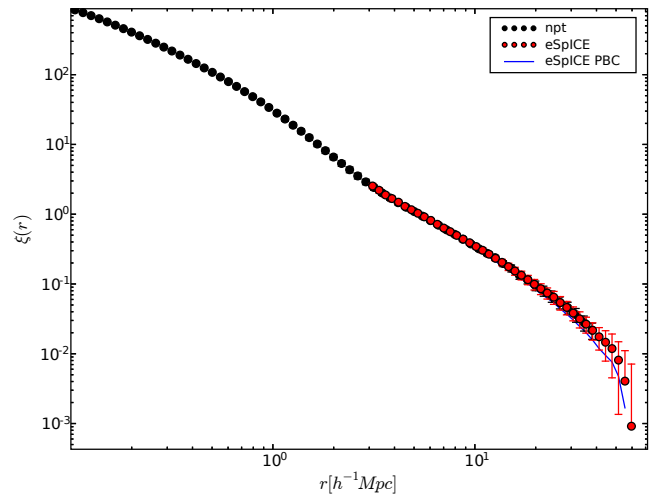}
\caption{The two-point correlation function measured
with the \cite{MooreEtal2001} tree-based algorithm 
(black symbols with errorbars) vs. our Fourier based approach (red/gray)
in a set of VLS simulations (see description in text).
The agreement is virtually perfect, despite the difference
in the methods and binning. For comparison, we also
show a separate measurement of the two-point function where (erroneously)
periodic boundary conditions were assumed (solid line). This
can be considered as a measurement without proper edge correction.
}
\label{fig:xi}
\end{figure}

The above algorithm to measure the correlation
function was implemented in {\tt C} using the {\tt FFTW3}
package for two and three dimensions.
The two-dimensional version of the algorithm was extensively
tested in \cite{BudavariEtal2003}, the first application to measure
the angular correlation function in the SDSS. In what follows,
we focus on three-dimensions.

To test our algorithm for the two-point correlation function,
and the power spectrum, 
we performed measurements in $\Lambda$CDM
simulations by the Virgo Supercomputing Consortium \cite{JenkinsEtal1998}.
We  used outputs of  VLS (Very Large Simulation)
with the following cosmological parameters: $\Omega_m=0.3$,
$\Omega_v=0.7$, $\Gamma=0.21$, $h=0.7$ and
$\sigma_8=0.9$. In order to estimate measurement errors,
we divided the VLS simulation into
eight independent subsets, thus each subset was $(239.5\mpc)^3$
cube.  Note that thus sub-cubes are {\em not} subject to periodic boundary
conditions, thus, although their geometry is simpler than a typical
galaxy survey, they are subject to edge effects, unlike a full
simulation cube.

We have measured the correlation function in logarithmic
bins in the range of $0.3-162 \mpc$
with our implementation of the algorithm {\tt eSpICE} (Euclidean
version of SpICE, the spatially inhomogeneous correlation estimator,
\cite{SzapudiEtal2001b})
Measurement of the correlation function on a $768^3$ grid takes 
about 15 minutes on one Opteron processor.
In addition, we also performed control measurements with
the \cite{MooreEtal2001} tree-based algorithm in logarithmic
bins from $0.1-38\mpc$ scales. The required CPU time of this latter
algorithm increases drastically on large scales, hence we
stopped at smaller scales. Even though the binning of the two runs
were not identical, the two algorithms produced virtually
identical curves as shown on Figure~\ref{fig:xi}. In addition,
we have measured the correlation function assuming --erroneously,
according to our previous discussion-- periodic boundary conditions.
The difference between the results based on periodic and non-periodic
boundary conditions demonstrate that edge effects are already significant
on $40\mpc$ scales in a $239.5\mpc$ size simulation.

The next step was calculating the edge corrected power spectrum.
First, we have performed a cubic spline interpolation on the
measured correlation function, to turn the discrete points into
a function, which can be sampled at the roots of Bessel functions.
Then we applied the recipe of Equation~\ref{eq:bessel} for $\nu=1/2$
to integrate the first of Eqs.~\ref{eq:2ptpk}. We used
$h = 1/32$ and truncated the sum at $N=200$; doubling
these values made no significant effect on our results. We have checked
the accuracy or our integration routine 
with $N=10000$ point integration according to
the direct double exponential integration by \cite{OouraMori1999},
and with {\tt Mathematica}. This showed that this fast 200 point
integration is better than $0.1\%$ accurate, which was good enough
for our purposes. Note that with some care,
{\tt FFTLog} \citep{Hamilton2000} could probably adapted to the 
same purpose, although the aliasing and numerical effects are
more delicate.

\begin{figure}[htb]
\epsscale{1.35}
\plotone{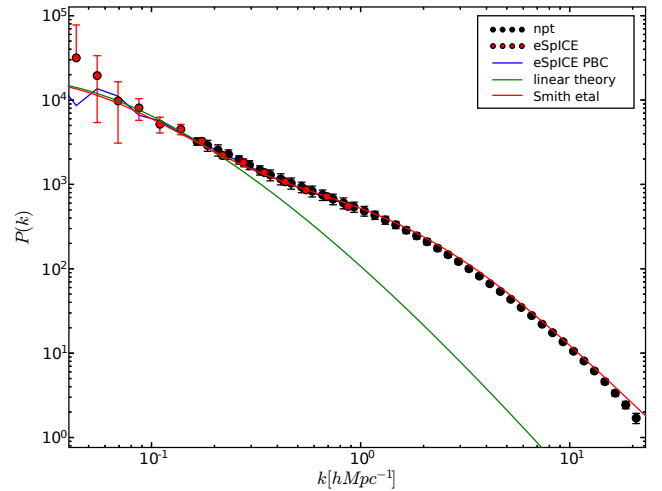}
\caption{The power spectrum extracted from the two-point correlation function using the method described in the
text. The green and red lines show the linear 
and non-linear \cite{SmithEtal2003} power spectra. The other symbols
and lines correspond to Figure~\ref{fig:xi}.
}
\label{fig:pk}
\end{figure}

The results are shown on Figure~\ref{fig:pk}. We have applied
the inversion both to the results from {\tt eSpICE}, and from
the tree based algorithm. We used a scale range conservatively
as $(k_{\min},k_{\max}) = {2\pi/r_{\max},1/r_{\min}}$. In addition
we plotted the linear theory and non-linear \cite{SmithEtal2003}
power spectra. The agreement of our reconstruction with the
phenomenology is nearly perfect.

\section{Summary and Discussions}

We have presented a fast Fourier based algorithm, which
is complementary to the tree-based algorithm of \cite{MooreEtal2001}.
While the latter is ideal for measuring the correlation function
on small scales, taking advantage of a KD-tree construction,
our grid based approach is ideal for large scales. Our algorithm
has the advantage that it does not slow down on large scales,
although the grid has a smoothing effect on the correlation function
on scales a few times the grid spacing. 

An interesting side effect of our algorithm is that as an intermediate
step it produces inhomogeneous correlation functinos of vector shifts 
$\xi({\bf r})$. These has been used in \cite{BudavariEtal2003}
to filter out systematic effects from the drift scanning operations
of the camera. In three dimensions, this could be used for
estimation of the redhisft space correlation function in the
distant observer approximation, for narrow, deep redshift surveys.

We also presented a method to turn a measured correlation function
into an edge corrected power spectrum. This is, apart from
numerical details, a direct Euclidean generalization of
the spherical method by \cite{SzapudiEtal2001b}. This new technique
is independent on the details of how the two-point
function is estimated: our Fourier based approach, a tree-based algorithm,
or even a direct $N^2$ algorithm can be used to measure the
two-point function. The inversion for the power spectrum should
work just the same. In the present work we did not correct
for grid effects in the inverted power spectrum, but it
should scale as $W(k)^2 = sinc(k)^{2D}$ in $D$ dimensions.

Were we measuring
the correlation function in linear scales (a built in option for {\tt eSpICE}
which does not influence the speed), we would have obtained
$P(k)$ in linear bins. The maximum $r_{\max}$ at which the correlation
function is measured, determines the resolution, and the smallest
$k_{\min} \simeq \Delta k \simeq 1/r_{\max}$ for $P(k)$, 
while the resolution and the smallest $r_{\min}$ at
which the correlation function is measured, determine 
$k_{\max} \simeq 1/r_{\min}$.

I accordance with the overwhelming majority of practical measurements
in the past, we choose to measure the correlation function in 
logarithmic bins. As a result we obtained $P(k)$ in 
approximately logarithmic bins in $k$. To see this, one
has to note that a fixed resolution corresponding to
a particular $k$ in the power spectrum, corresponds to a maximum $r$
at which the correlation function is measured at this
resolution.  This maximum  translates into
an effective resolution $\Delta k$ at this particular $k$.
This qualitative argument can be made more rigorous
with calculation of the effective window function if needed.

Large scale structure studies, up until now, 
have estimated ``pseudo power spectra''
(a term borrowed from cosmic microwave background research)
i.e. one convolved with the survey geometry.
The potential of detecting baryonic oscillations in the
measurement of the power spectrum motivates the need
for measuring power spectra with the highest resolution
possible. In particular, if one of the survey dimensions
is much smaller than the other two, the effective convolution
kernel will be much wider than desirable. In the above
we demonstrated that the power spectrum can be recovered
from steps not any more complicated than the measurement
of the corresponding pseudo power spectrum. While the
information content cannot be improved with the inversion
\citep[e.g.][]{Efstathiou2004}, it is immensely useful
when comparing measurements from different surveys, from
different geometries, as it has been common in
CMB research. 

Note that the present technique can be used to estimate
$C_\ell$'s in the flat sky approximation, using the asymptotic
behavior of the Legendre polynomials 
$P_\ell(\cos\theta)\simeq 
\sqrt{\frac{2}{\pi\ell\sin\theta }} \cos((l+1)\theta-\pi/4)$,
and the Bessel functions 
$J_0(z) \simeq \sqrt{\frac{2}{\pi z}}cos(z-\pi/4)$.
The correlation function can be measured with {\tt eSpICE},
and integration of Equation~\ref{eq:2ptpk} gives the $C_\ell$'s
with approximately $k \simeq \ell+1/2$.
Generalization of the proposed computational and inversion
techniques to gravitational lensing follows directly
from the generalization {\tt SpICE} to CMB polarization,
such as \cite{ChonEtal2004}. Finally, generalization of
Equation~\ref{eq:2ptpk} for a relation between the three-point
function and the bispectrum \citep{Szapudi2004a} together
with a fast algorithm to measure the three-point function
yields a new, edge corrected method to measure the bispectrum.
These generalizations will be published elsewhere.

Our {\tt eSpICE} implementation will be made public upon acceptance
of this paper.

I thank Alex Szalay for stimulating discussions, Josh Hoblitt
and Gang Chen for computing help.
IS and JP was supported by NASA through AISR NAG5-11996, 
and ATP NASA NAG5-12101 as well as by
NSF grants AST02-06243, AST-0434413 and ITR 1120201-128440.
JP acknowledges support of  PPARC Grant number
PPA/G/S/2000/00057.

%%%%%%%%%%%%%%%%%%%%%%%%%%%%%%%%%%%%%%%%%%%%%%%%%%%%%%%%%%%%%
%%%%% References %%%%%

% to use bibtex files
%\bibliography{bcgs,szapudi2004,bl,nsf04,mybib} 
%\bibliographystyle{apj}  %>>>> makes bibtex use spiebib.bst
%%%%%%%%%%%%%%%%%%%%%%%%%%%%%%%%%%%%%%%%%%%%%%%%%%%%%%%%%%%%%

%to include bibliography

%%%%%%%%%%%%%%%%%%%%%%%%%%%%%%%%%%%%%%%%%%%%%%%%%%%%%%%%%%%%%

\end{document}